\documentclass[final,5p,times,twocolumn]{elsarticle}

\usepackage{amssymb}
\usepackage{lipsum}
\usepackage{amsmath}
\usepackage{xcolor}

\usepackage{hyperref}

\journal{Astronomy $\&$ Computing}

\begin{document}

\begin{frontmatter}


\title{Estimation of Orbital Parameters from $(u,v)$-coverage for a Space Radio Interferometer}

\author[first,second]{I. I. Bulygin}
\author[first]{M. A. Shchurov}
\author[first]{A. G. Rudnitskiy}
\affiliation[first]{organization={Astro Space Center, Lebedev Physical Institute, Russian Academy of Sciences},%Department and Organization
            addressline={Profsoyuznaya st., 84/32}, 
            city={Moscow},
            country={Russia}}
\affiliation[second]{organization={Sternberg Astronomical Institute of Lomonosov Moscow State University},%Department and Organization
            addressline={Universitetsky pr., 13}, 
            city={Moscow},
            country={Russia}}           

\begin{abstract}
Finding a suitable very long baseline (VLBI) interferometer geometry is a key task in planning observations, especially imaging sessions. The main characteristic of the quality of VLBI imaging data is the $(u,v)$-coverage. In the case when one or more radio telescopes are located in space, this task becomes more complex. This paper presents a method for recovering the optimal orbital parameters of space radio telescopes for a given desired $(u,v)$-coverage, which in turn is the inverse task of searching for the optimal geometry and orbital configurations of space-ground and pure space VLBI interferometers.
\end{abstract}




\end{frontmatter}

%\tableofcontents

%\linenumbers

\section{Introduction}
\label{sect:intro}
Very long baseline interferometry (VLBI) is a powerful technique for obtaining detailed images of astrophysical sources. Unlike other astronomical methods, VLBI allows for angular resolutions of tens of microarcseconds, which cannot be achieved by any other methods in the modern astronomy. Progress in the development of ground-based VLBI systems in the centimeter range, like, for example, the Very Long Baseline Array (VLBA), European VLBI Network (EVN) was followed by the implementation of space-ground interferometers \citep{Lovell1999,Ulvestad2000,Mochizuki2003,Kardashev2013}. Highly Advanced Laboratory for Communications and Astronomy (HALCA) or VSOP (VLBI Space Observatory Program) \citep{Lovell1999} and Radioastron were successful among them \citep{Kardashev2013}. The second one, having the shortest operating wavelength of 1.3~cm, achieved a still unbeaten record in angular resolution of 8~$\mu$as \citep{Baan2022}.

Over time, interest of astronomers shifted towards millimeter wavelengths. This opened up a new era in high-resolution radio astronomy. Successful imaging observations of the SMBH M87 and Sgr~A* with the Event Horizon telescope \citep{EHT2019a,EHT2022a} marked the further rapid development towards the millimeter VLBI \citep{Palumbo2018, Raymond2021, Ricarte2023}.

Current trends take into account the experience gained in ground millimeter VLBI and concentrate on the development of pure space-to-space high-frequency VLBI instruments \citep{Hong2014,Gurvits2020}. The fact is that Earth limits VLBI capabilities in terms of maximum achievable angular resolution. In addition, it limits capabilities in terms of the opacity of the atmosphere in the high-frequency range. The last few years have marked the emergence of a number of such concepts: EHI \citep{Roelofs2019,Fish2020}, THEZA \citep{Gurvits2021}, CAPELLA \citep{Trippe2023}, and proposals for optimal orbital configurations for such systems have been analyzed \citep{Rudnitskiy2023}. It is the optimal geometry, both in the case of ground, space-ground, and especially pure space VLBI instruments, that plays a key role in the correct, efficient and successful planning and conducting of VLBI observations.

To obtain highly detailed VLBI images, optimization of the $(u,v)$-coverage, which is directly related to the resulting quality of the resulting image, plays a critical role \citep{Zhang2021,Liu2021}. In this regard, the task of searching for optimal geometry for various, primarily space, VLBI instruments becomes urgent. Usually such tasks are solved by searching for the VLBI geometry/orbit parameters to obtain the most suitable $(u,v)$-coverage. This paper presents and discusses a different method for searching for the optimal geometry of space and space-ground VLBI interferometers by solving the inverse problem of reconstructing space radio telescope orbital parameters from a desired $(u,v)$-coverage\footnote{Source code of developed software is available: \href{https://github.com/bullygen/OrbitEstimator_fromUV}{github.com/bullygen/OrbitEstimator\_fromUV}}.

\section{Direct problem: Calculation of $(u,v)$-coverage}
\label{sect:direct}
In very long baseline radio interferometry (VLBI), obtaining a high-quality image of sources using aperture synthesis methods directly depends on the quality of the $(u,v)$ coverage. To begin with, let us consider in more detail how the Cartesian coordinates $x, y, z$ of the telescope are related to the $(u,v)$ coordinates.

If $\mathbf{r} = (x, y, z)$ is the base vector $\vec{B}$ between two telescopes in the ICRF system, and $(\alpha, \delta)$ -- J2000 coordinates of the source taking into account aberration, then $(u,v)$ coordinates are obtained according to the following procedure:

\begin{equation}
    \begin{split}
        u &= - x \sin \alpha + y \cos \alpha \\
        v &= - x \cos \alpha \sin \delta - y \sin \alpha \sin \delta  + z \sin \delta \\
        w &= x \cos \alpha \cos \delta + y \sin \alpha \cos \delta  + z \cos \delta
    \end{split}
    \label{eq:uvw}
\end{equation}

Each moment of time $t_i, i = 1,\dots,N$, when a signal is received during the observation time interval $t_i \in [0, T]$, it is necessary to know the coordinates in the ICRF system of telescopes (both space and ground-based) $ \mathbf{R}_n$ to calculate the base vector between them:

\begin{equation}
    \mathbf{r}_{nm}(t_i) = \mathbf{R}_n(t_i) - \mathbf{R}_m(t_i)
\end{equation}

Here the indices are $n \neq m$; $n,m = 1,\dots,M$ run through the entire set of telescopes that participate in the observations. This gives $M(M-1)/2$ independent base vectors.

As can be seen from (\ref{eq:uvw}), if we exclude the third equation responsible for $w$-coordinate (as usually happens when reconstructing VLBI images), then for a given $(u,v)$ coverage, it will be impossible to restore the original $\mathbf{r} = (x, y, z)$. However, it will be possible to talk about a certain family of orbits that make it feasible to obtain the original (or as close as possible to it) $(u,v)$ for the selected telescopes during the considered time period. The solution to this problem is discussed below.

Space telescope positions are calculated based on a constant elliptical spacecraft orbit. Ground telescope positions are calculated by relying on the theory of Earth rotation \cite{IAU_nut_model} and the conversion of coordinates from ITRF to ICRF.

Before calculating the coordinates $(u, v)_{nm}(t_i)$ in the plane perpendicular to the direction of the source, it is necessary to verify its potential observability with each telescope at this moment in time. To do this, it is checked that for a ground telescope the source is above the horizon, and for a space telescope it is not located behind the Earth and its angular distance to the Sun is greater than a certain threshold $\theta_\odot$, which must be set based on information about the constraints of the telescope and its instruments.

\section{Inverse Problem: Obtaining Orbital Parameters from $(u,v)$ coverage}
\label{sect:inverse}

Let us denote the input pixelated $(u,v)$ coverages as $I_{ij}^{\,(k, 0)}$ for each source $k$, the array of ground telescopes and source coordinates as $\mathbf{Y} $. The process of finding the optimal orbit of telescopes in the space VLBI (SVLBI) is the following:

\begin{enumerate}
\item The initial approximation $\mathbf{X}_0$ is calculated for the satellite constellation orbital parameters.

As an initial approximation for one space observatory, the telescope orbit is located in a plane perpendicular to the source direction. The dimensions and eccentricity of the initial approximation orbit can be estimated from the maximum and minimum base during the observation time of the source $\mathbf{r}_\text{min}$ and $\mathbf{r}_\text{max}$:
    
\begin{equation}
    \begin{split}
        e_0 &= \dfrac{r_\text{max} - r_\text{min} - 2R_\odot}{r_\text{min} + r_\text{max}} \\
        a_0 &= \dfrac{r_\text{min} + r_\text{max}}{2} \\
        i_0 &= 90^\circ - \delta \\
        \Omega_0 &= 90^\circ + \alpha \\
        \omega_0 &= \mathtt{arctan2}(u_\text{min}, v_\text{min}) \\
    \end{split}
\end{equation}

Here, the function $\mathtt{arctan2}(y, x)$ has a range $[0, 2\pi)$ and gives an angle based on two components of a vector $(x, y)$. Selecting an initial approximation for multiple space telescopes that observe multiple sources is a task for future studies.
    
\item Functionality is generated to minimize the following forms:

\begin{equation}
     \begin{split}
        &\mathcal{L}(\mathbf{X}\, |\, \mathbf{Y}) = \\
        & =\sum_{k\, \in\, \text{sources}}\sum_{(i, j) \, \in \, \text{ray}_k} \left(\text{RAY}\left[\left(I^{\,(k)}(\mathbf{X}\, |\, \mathbf{Y}) - I^{\,(k, 0)}\right) * G(\sigma) \right]_{ij}\right)^2 
    \end{split}
    \label{eq:fuctional}
\end{equation}

where $\mathbf{X}$ is a vector of parameters used for minimization. These are either Keplerian orbital elements or the coordinates of the vectors of the Kholshevnikov metric (see~\ref{sect:appA}), although in the general case it is acceptable to use another metric for comparison. $\mathbf{Y}$ is an array of constant parameters, namely: ITRF coordinates of ground telescopes at a given time moment, source coordinates for the J2000 epoch, observation time points and physical parameters involved in the conversion of coordinates between the ICRF and ITRF systems. RAY -- synthesized beam of VLBI interferometer, which is a direct Fourier transform of the (u,v) coverage with normalization by 1.
    
Convolution of the difference between $(u,v)$ coverage and a Gaussian kernel of a certain characteristic size $\sigma$ is necessary so that similar patterns in the recommended and model coverage attract each other, forming a similar beam during the minimization process. It must be done after the Fourier transform of the coverage, then replaced by pixel-by-pixel multiplication.

It is the difference between beams that is minimized, not the coverage, because the coverage is calculated from a discrete set of points, which means it is necessary to convert it into some kind of set of pixels, in which the observing time of the source in each pixel must be taken into account \cite{SVLBI_optim_1}. However, in \cite{SVLBI_optim_2} a different approach was used -- the side lobes of the beam and the eccentricity of its peak were minimized. Ideologically, the method described in this work is a generalization of the method from \cite{SVLBI_optim_2} with the difference that in the current work the beam is compared with a beam already predetermined for each source, taking into account ground telescopes.

It should be noted that it is the beam, depending on the $(u,v)$ coverage, that is used to form the image of the source, so its shape is of primary importance to the observer. However, we note that, by virtue of Parseval’s theorem, this functional is proportional to the sum of squared deviations of one pixelated coverage from another:

\begin{equation}
    \sum_{(i, j) \, \in \, \text{ray}} \left|\text{FFT}\left[I\right]_{ij}\right|^2 \propto \sum_{(i, j) \, \in \, \text{UV-coverage}} \left|I_{ij}\right|^2
\end{equation}

This paper solves the problem of orbit design based on coverage characteristics. For observations, not the entire frequency range of the initial coverage (or beam) is of importance, so the convolution of the coverage with a Gaussian kernel with a characteristic spatial scale $\sigma$ is introduced into the functional. In frequency space, this becomes the multiplication of the beam by a Gaussian kernel with a characteristic angular scale of $\lambda/\sigma$. This is a low-pass filter of the source data, which allows you to select different beam reconstruction accuracy.

\item The functional is minimized (\ref{eq:fuctional}). The value of the functional strongly depends on small fluctuations in the orbital parameters (for example, tracks from different orbits will overlap each other in the $(u,v)$ plane, generating the same projections of the bases), so a large number of local minima are expected. It is because of this that the Powell method was used to minimize the functional, as it requires fewer calculations than gradient methods \cite{NelderMead}.

\end{enumerate}

\section{Implementation}
\label{sect:implementation}

To build the coverage, three objects are required with corresponding classes that contain methods for calculating their parameters:

\begin{enumerate}
     \item Class \texttt{Source} contains fields of equatorial coordinates for the J2000 epoch, coordinate translations between different epochs and reference systems, including relativistic aberration effects. The proximity of the source to the Sun is taken into account. This makes it possible to consider the minimum angular distance to the Sun as a source of illumination.
     \item Class \texttt{GroundTelescope} performs similar tasks for a given array of ground telescopes, checks the visibility of a source above the horizon. When converting coordinates from ITRF to ICRF, IERS Earth rotation parameters forecasts (IERS Bulletin A) for a specific date are used.
     \item Class \texttt{Satellite} calculates the trajectory of a space telescope over a given period of time using either classical celestial mechanics, analytically taking into account secular perturbations of orbital elements, or numerically. The possibility of the source being eclipsed by the Sun is considered.
\end{enumerate}

Arrays of source objects, ground telescopes and space telescopes are created, and the observation time for each source is specified. After this, the set of base vectors $\mathbf{r}$ is modeled and converted into $(u,v,w)$ coordinates for each source (with coordinates $\alpha, \delta$):

The base vectors are written into a pixelated image of a size convenient for FFT (a multiple of a power of two, as the most optimal). If several observation points fall on one pixel, its weight is equal to the number of observations on a given base.

All code for modeling and solving the inverse problem was implemented in the \texttt{Python} language using the \texttt{numpy}, \texttt{matplotlib}, \texttt{scipy} and \texttt{pysofa} libraries.

\section{Testing Procedure}
\label{sect:test1}

Both real and synthetic data were used to test the orbit design algorithm's effectiveness:

\begin{itemize}
     \item Dataset 1. 10 artificially generated $(u,v)$ coverages for 1 satellite and 1 source according to the algorithm described in the Section~\ref{sect:direct} of size $64 \times 64$ pixels. In addition, the following were transmitted separately:
     \begin{itemize}
         \item coordinates of the ground telescope $\mathbf{r}_\text{ITRF}$;
         \item J2000 source coordinates $\alpha, \delta$;
         \item scale -- 1 pixel in Earth diameters;
         \item the minimum permissible distance between the source and the Sun (in angular measure $\theta_\odot = 45^\circ$);
         \item duration of observations -- 3 solar days;
         \item Julian date of observations beginning. JD$=2460473.5$ was selected (June 12th, 2024.
         \item number of observation points $N$. Everywhere $N = 3000$.
     \end{itemize}

     \item Dataset 2. Real observations of the Radioastron mission and 13 ground-based telescopes in the UTC time interval from 04/04/2014 11:00:00 to 04/05/2014 05:00:00, converted into a format for synthetic orbits. The source 0851+202 (OJ287) was observed at a wavelength of $\lambda = 1.36$~cm.

     \item Dataset 3. Artificially generated $(u,v)$ coverages for one plane of orbit with two satellites of the EHI project \cite{Kudriashov2021a} in the direction to Sgr~A*, the observing time and other observation parameters coincide with Dataset 1. An algorithm with one target source and one satellite was used to restore the orbit. This test is necessary to fundamentally check whether, using an algorithm, it is possible to design the orbit of one satellite with similar quality, and in less signal accumulation time.
\end{itemize}

For each test, two variants of the parameter space $\mathbf{X}$ were used to study the efficiency: Keplerian orbital elements $\mathbf{X}_\text{Kep} = (a, e, i, \Omega, \omega, M_0 )$ and components of vectors $\mathbf{u}$, $\mathbf{v}$ of the Kholshevnikov metric: $\mathbf{X}_\text{Khol} = (u_x, u_y, u_z, v_1, v_2, M_0)$ (see \ref{sect:appA} for details). It is possible to compare the effectiveness of two methods according to several parameters $q_i$:

\begin{enumerate}
     \item Closeness of the final fitting result ($f$) to the true one ($r$) according to the Kholshevnikov metric:
     \begin{equation}
         q_1 = \frac{d_{fr}}{\sqrt{a/D_\oplus}}
     \end{equation}
     It shows how the designed orbit differs from the synthetic one (which is not necessarily optimal).

     \item Normalized finite ($f$) minimization functional:

     \begin{equation}
         q_2 = \frac{\mathcal{L}(\mathbf{X}_f\, |\, \mathbf{Y})}{\mathcal{L}(\mathbf{X}_0\, |\, \mathbf{ Y})}
     \end{equation}

     It shows how much coverage has been optimized compared to the rough initial guess. The value of the functionality will be considered satisfactory in Datasets 2 and 3 if it is less than the average value for all tests in the data from point 1, presented in Table \ref{tab:results}.
\end{enumerate}

\section{Results}
\label{sect:test2}
\begin{figure*}[h]
    \centering
    \includegraphics[width = 0.8\textwidth]{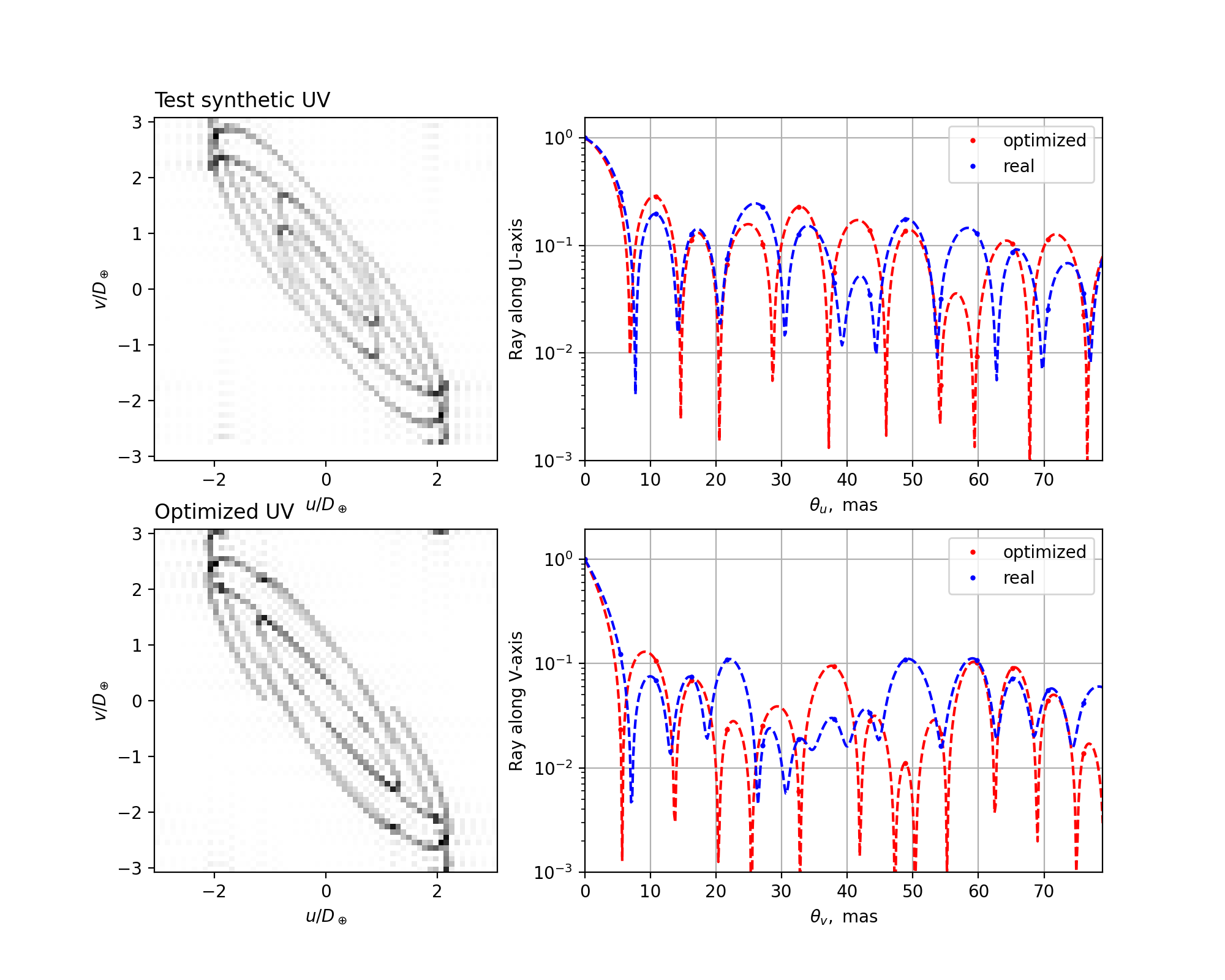}
    \caption{Dataset 1, Test 2. \textit{Top left} -- synthetic model of $(u,v)$ coverage, \textit{Bottom left} -- reconstructed $(u,v)$ coverage, on both axes dimensions are expressed in Earth diameters (without reference to wavelength). \textit{Right} -- comparison of test synthetic and reconstructed beams along the $u$ and $v$ axes on a logarithmic scale. The X-axis shows the angular size up to $3\lambda/\sigma$ at the test wavelength $\lambda = 1$ m. The ordinate is the normalized beam intensity. There is a coincidence of the general trend at a typical frequency filter width -- $\lambda/\sigma$.}
    \label{fig:test_02}
\end{figure*}

\begin{figure*}[h]
    \centering
    \includegraphics[width = 0.8\textwidth]{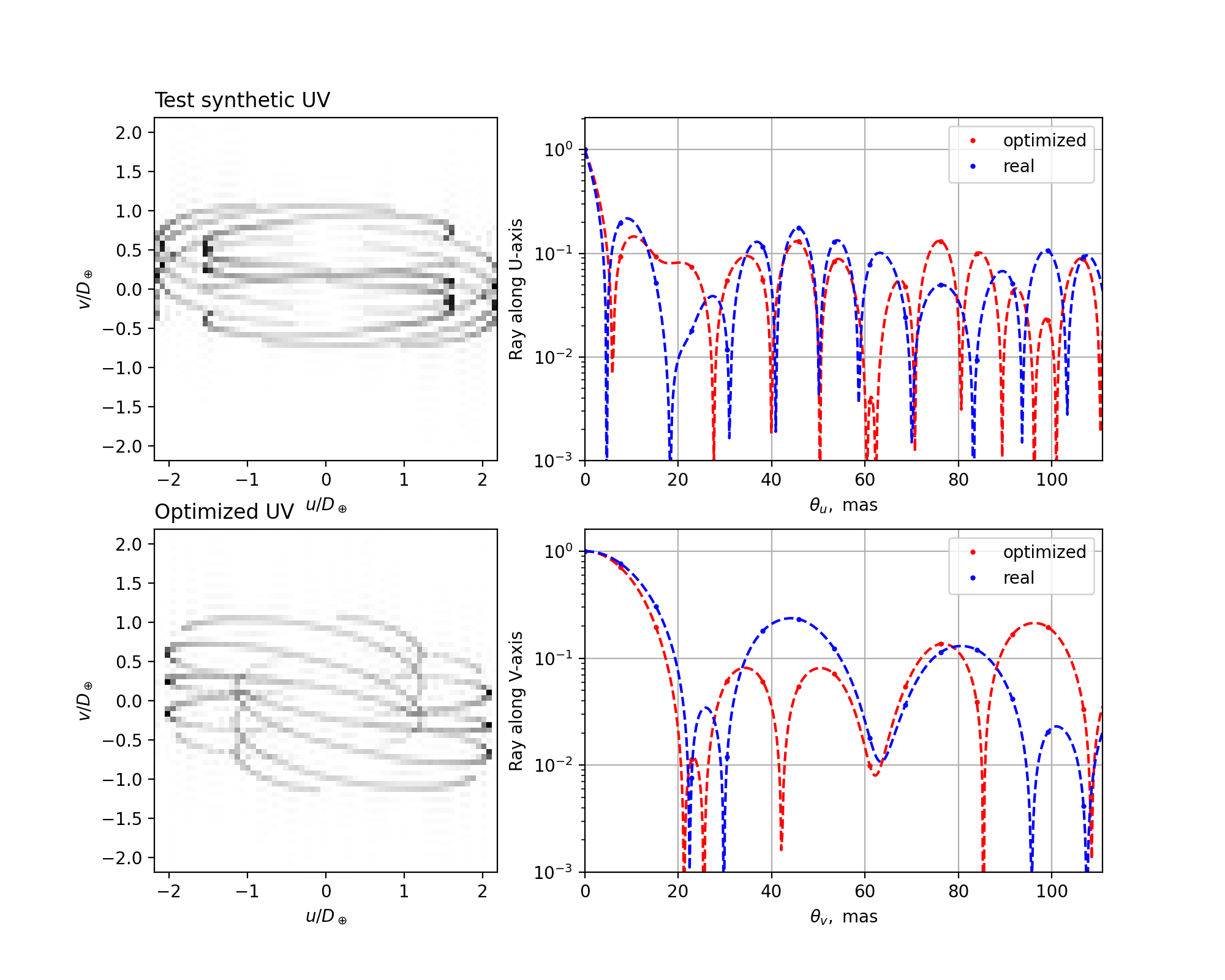}
    \caption{Dataset 1, Test 5. The description is same as for the Test 2 (Fig.~\ref{fig:test_02}).}
    \label{fig:test_05}
\end{figure*}

\begin{figure*}[h]
    \centering
    \includegraphics[width = 0.8\textwidth]{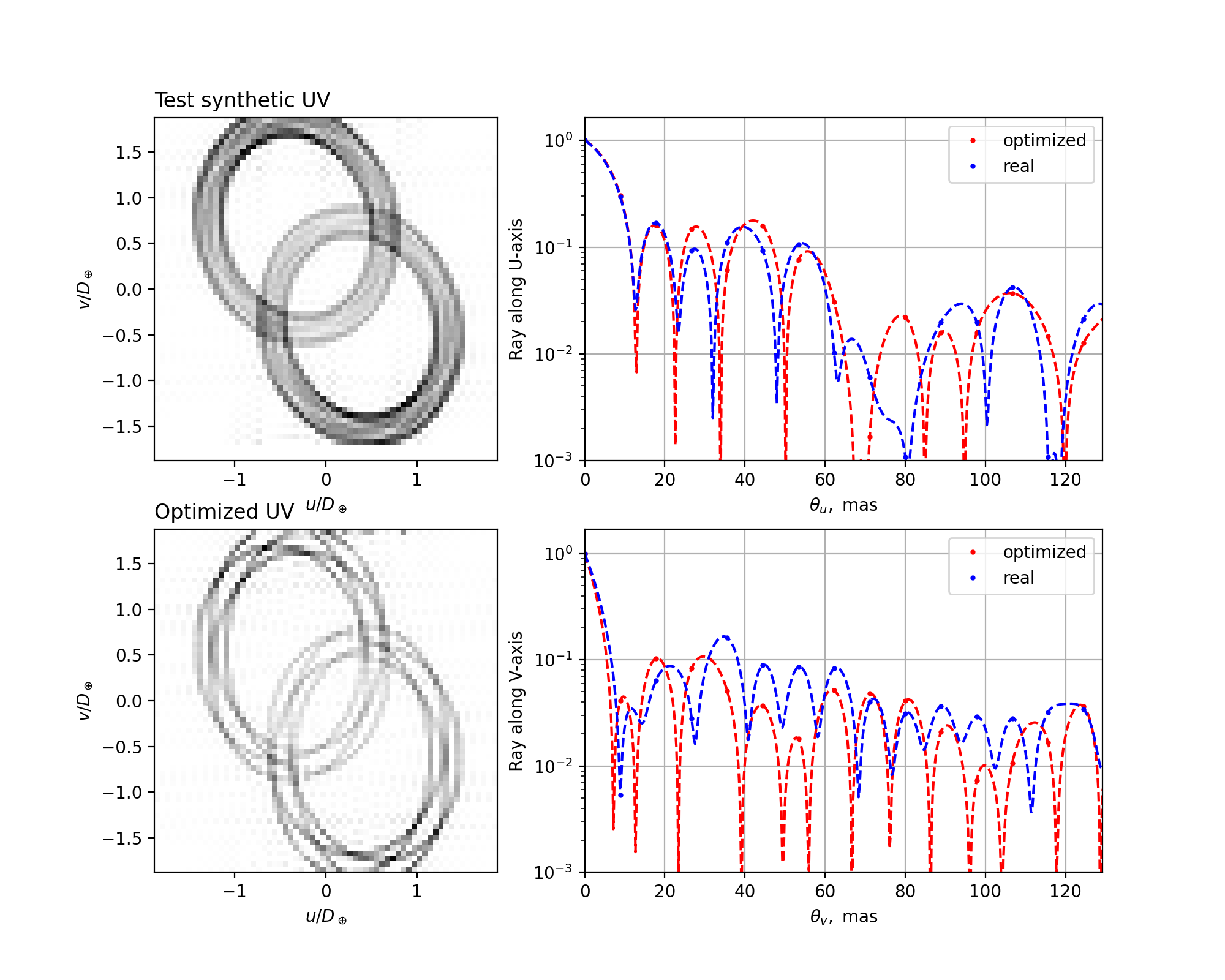}
    \caption{Dataset 1, Test 6. The description is same as for the Test 2 (Fig.~\ref{fig:test_02}).}
    \label{fig:test_06}
\end{figure*}

Testing results on Dataset 1 are presented in Fig. \ref{fig:test_02}, \ref{fig:test_05}, \ref{fig:test_06}. Dataset 1 shows the results using Keplerian elements as variables for 3 of the 10 tests.

After all tests were carried out, the averages $q_1$ and $q_2$ were estimated. The results are presented in Table \ref{tab:results}.

\begin{table}[h]
    \centering
    \begin{tabular}{|c|c|c|}
        \hline
              & Kholshevnikov Elements  & Kepler Elements \\ \hline
        $q_1$ & $1.14 \pm 0.43$         & $1.23 \pm 0.42$ \\ \hline
        $q_2$ & $0.45 \pm 0.22$         & $0.26 \pm 0.17$ \\ \hline
    \end{tabular}
    \caption{Results for Dataset 1, averaged over all tests.}
    \label{tab:results}
\end{table}

According to these indicators, the functional for minimization on average decreases by a significant number of times ($q_2$) when using Kepler elements as variables, so they are recommended for further use in the design process. Kepler elements will also be used in the design based on Datasets 2 and 3. Interestingly, the test orbit, obviously not the optimal one, is removed by the Holshevnikov metric ($q_1$) from the designed orbit when considering any set of variables. Thus, we can assume that a configuration has been found that is similar to the given one in terms of the quality of observations of the selected source. However, it is different from the original one. By adding additional multipliers to the functionality that are responsible for other technical constraints of the SVLBI project (for example, such as fuel economy, observation time, etc.), it is possible to find an orbit with similar image quality, but satisfying additional criteria.

\begin{figure*}[h]
    \centering
    \includegraphics[width = 0.8\textwidth]{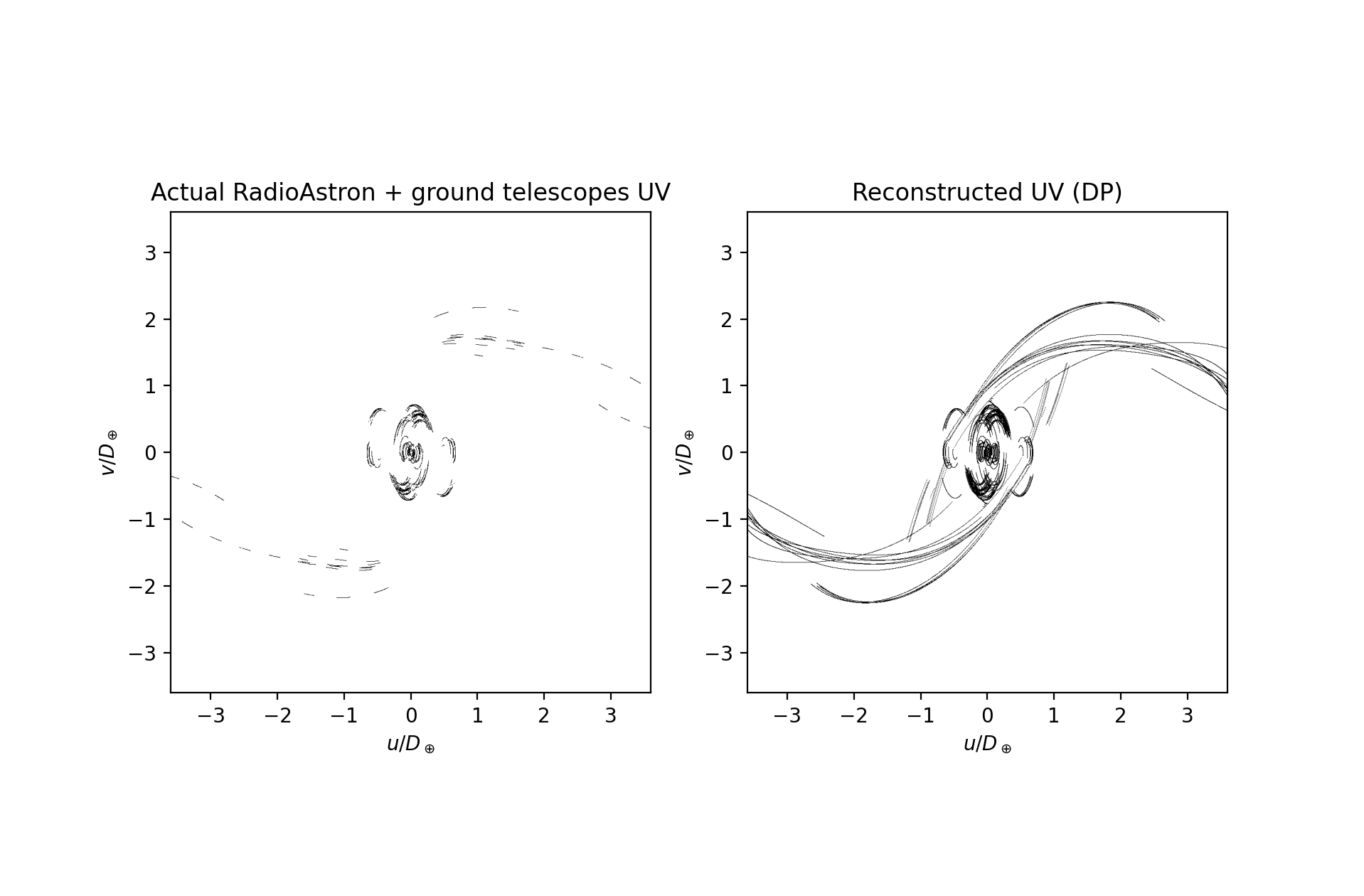}
    \caption{$(u,v)$ coverages for Dataset 2. On the left -- the real $(u,v)$ coverage, on the right -- the restored one.}
    \label{fig:RA_direct}
\end{figure*}

On Dataset 2 (Radioastron mission), both the design method and the orbit reconstruction (obtaining Keplerian elements) were tested. During the design, $(u,v)$-coverage was simulated for a system of one space and 13 ground telescopes with known orbital elements and coordinates, respectively. The results are presented in Fig. \ref{fig:RA_direct}.

It is noticeable that recovery performed with fairly high accuracy. However, several artifacts distinguish the real coverage from the simulated one. They are caused by three main factors:

\begin{itemize}
     \item Instrumental limitations of the space telescope (periodic cooling of the high gain antenna, visibility conditions from tracking stations, etc.);
     \item Low quality of correlation between some telescopes, due to which observations at this moment are not subject to joint synthesis and for this reason are excluded from the $(u,v)$ coverage during the processing procedure;
     \item Limited observation time of some ground telescopes from the entire set.
\end{itemize}

This complicates Radioastron orbit reconstruction based on real coverage, because the optimization may collapse to the small bases for which ground telescopes are responsible. Several other facts also interfere, such as, for example, the non-optimally of the initial orbit. To simplify the orbit reconstruction for a specific mission, it was decided:

\begin{itemize}
     \item To use Keplerian orbital elements as variables, since research on test data gives better design quality for them;
     \item Due to the highly ellipticity of the spacecraft orbit, use the limitation on the minimum height of the satellite above the Earth’s surface:
     \begin{equation}
         h_\text{min} = a(1 - e) - R_\oplus > 600 \,\text{km}
     \end{equation}
     This limitation is implemented in the minimization method by cutting off the simplex if at some iteration it falls outside the boundaries of the admissible parameter space.
\end{itemize}

\begin{figure*}[h]
    \centering
    \includegraphics[width = 0.8\textwidth]{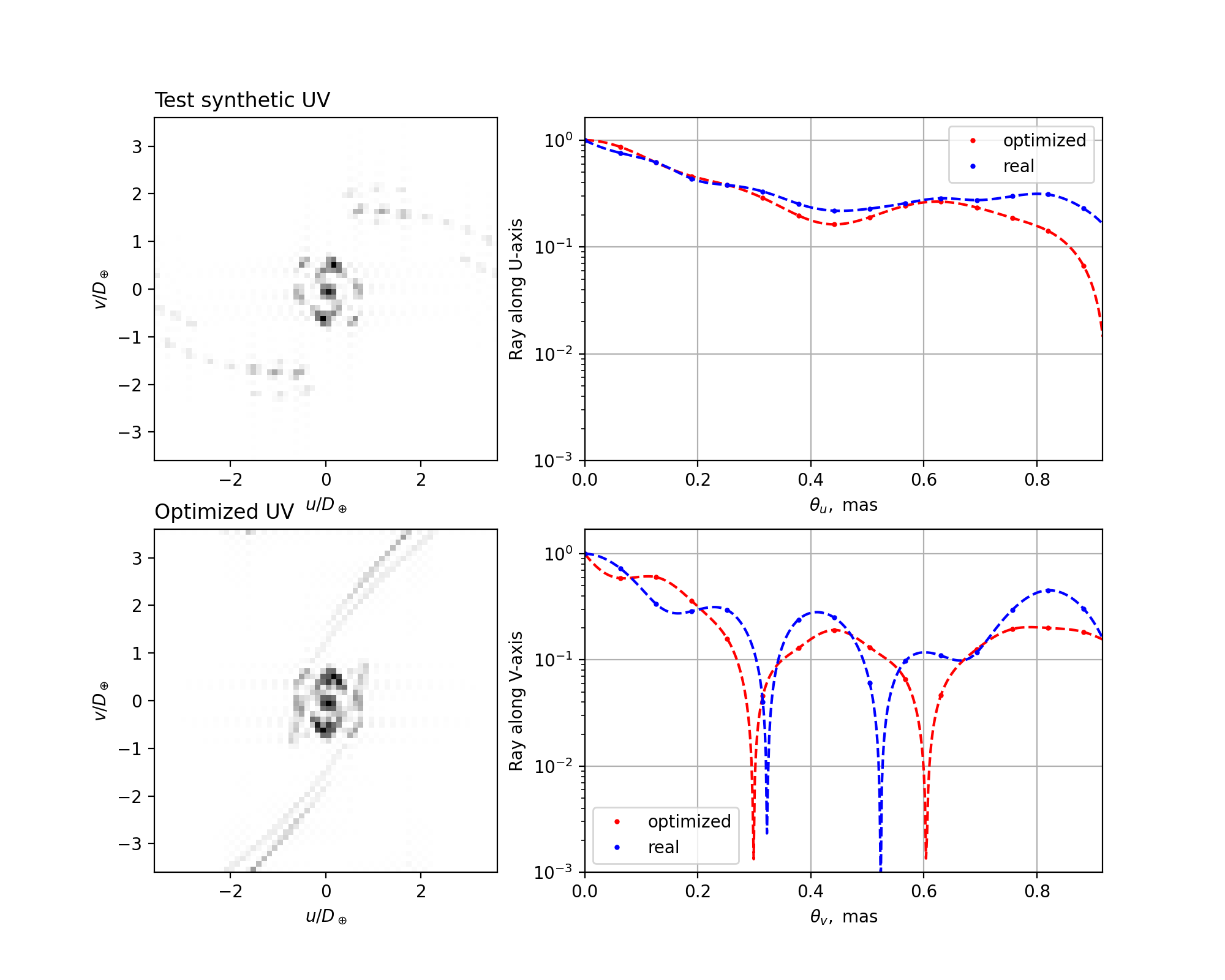}
    \caption{Dataset 3, an orbit reconstructed from a given occupancy, based on the actual occupancy of the RadioAstron mission observation session. The description is same as for the Test 2 (Fig.~\ref{fig:test_02}).}
    \label{fig:RA_fit}
\end{figure*}

Fig.~\ref{fig:RA_fit} presents the optimization results in a format similar to that described above, as well as in the Table \ref{tab:RA_params} with a comparison of real orbital elements for this date. For this design case, $q_2 = 0.11$, which is less than the average parameter value for synthetic coverages (Dataset 1), and therefore is a satisfactory result. As can be seen from the table, an accurate restoration of the orbital parameters in this case was not carried out. This is due to the fact that this observation session was carried out near the periapsis of the orbit, as well as the fact that the orbit itself is not optimal for this source, which ultimately led to a smaller semi-major axis and a different inclination.

\begin{table}[h]
    \centering
    \begin{tabular}{|c|c|c|}
        \hline
        Element             & Real      & Optimized \\ \hline
        $a$, $10^3$ km      & $175.14$  & $53.42$   \\ \hline
        $e$                 & $0.925$   & $0.841$   \\ \hline
        $i,\,^\circ$        & $20.1$    & $57.3$    \\ \hline
        $\Omega,\,^\circ$   & $128.3$   & $180.0$   \\ \hline
        $\omega,\,^\circ$   & $161.8$   & $-123.4$  \\ \hline
        $M_0,\,^\circ$      & $-30.5$   & $-0.3$    \\ \hline
    \end{tabular}
    \caption{The actual orbit of the Radioastron mission at the date of observation and the design result.}
    \label{tab:RA_params}
\end{table}

\begin{figure*}[h]
    \centering
    \includegraphics[width = 0.8\textwidth]{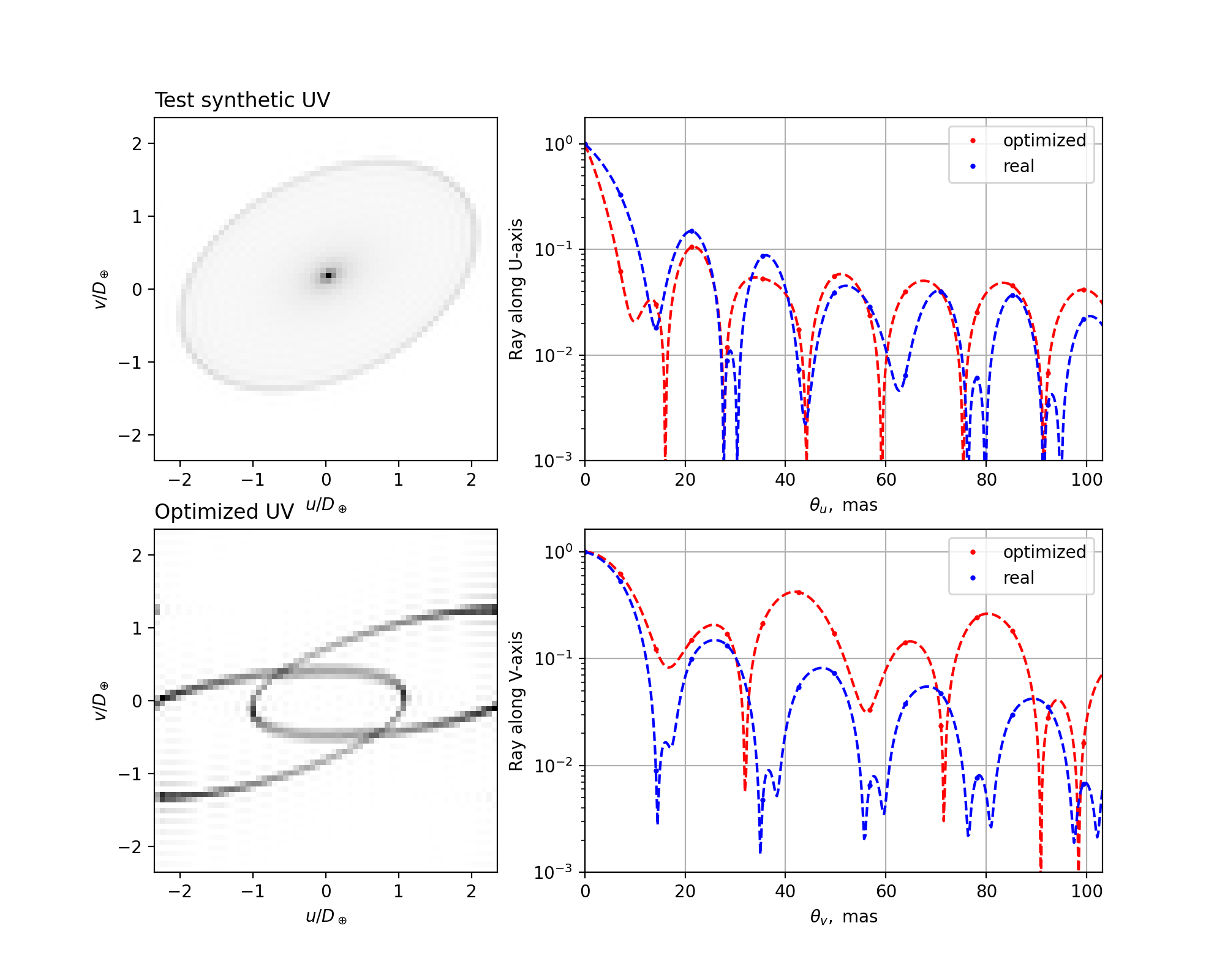}
    \caption{Result of designing the orbit of one satellite for the $(u,v)$ coverage of two space telescopes of the EHI mission.}
    \label{fig:EHI_fit}
\end{figure*}

As an additional test of the algorithm for one space satellite, the case with coverage generated by two space telescopes was considered. This was done following the EHI SVLBI project concept. In this case, the test was to give the algorithm an initially unreallizable, ``ideal'' coverage and see how it copes with this task. In fact, in this case, the problem of finding the most complete coverage with an interferometer consisting of one ground and one space telescope in three days was solved. The optimization result is shown in Fig. \ref{fig:EHI_fit}, with $q_2 = 0.22$, which indicates a satisfactory design result with respect to synthetic coverages (Dataset 1). Table~\ref{tab:EHI_params} gives Keplerian elements. 

\begin{table}[h]
    \centering
    \begin{tabular}{|c|c|}
        \hline
        Element             & EHI Optimized \\ \hline
        $a$, $10^3$ km      & $26.18$       \\ \hline
        $e$                 & $0.669$       \\ \hline
        $i,\,^\circ$        & $106.1$       \\ \hline
        $\Omega,\,^\circ$   & $107.8$       \\ \hline
        $\omega,\,^\circ$   & $-101.9$      \\ \hline
        $M_0,\,^\circ$      & $0.4$         \\ \hline
    \end{tabular}
    \caption{Keplerian elements for Dataset 3.}
    \label{tab:EHI_params}
\end{table}

As can be seen from the test results, an acceptable resolution was achieved. This corresponds to a baseline order of $26 \cdot 10^3$ km, which is provided by one satellite instead of two. On 2 cross-sections of the beam along the $u$ and $v$ axes it is also clear that the beam structure is coincide with the beam structure of the EHI mission in the two first maxima. The relative $(u, v)$ coverage density of the original mission design is concentrated at the edge of the coverage region, resulting in a completely analogous orbital pattern for a single satellite.

\section{Discussion and Conclusions}
\label{sect:conc}

A software was developed that allows, to reconstruct orbits of space telescope orbits for a given $(u,v)$-coverage. The results of the tests showed that reconstructed orbits make it possible (within the capabilities of constructing stable orbits) to implement both $(u,v)$ coverages and interferometer beam pattern that are as similar as possible to the original ones. This software is extremely helpful and it will be used for planning the scientific observation program of future SVLBI missions, such as, for example, the Millimetron space observatory.

During the software development, a solution to the problem of numerical pixel-by-pixel comparison of the $(u,v)$ coverage was proposed. Instead of comparing two pixelized coverages, it was proposed to solve the problem using a linear Fourier transform. This was done by comparing interferometer beam patterns and its cross section trends.

Due to the fact that the original problem is mathematically incorrect, i.e. the number of unknown parameters exceeds the number of equations, by minimizing the described above functional the algorithm finds stable orbits belonging to families that satisfy the initial requirements for $(u,v)$ coverage. It was shown that the found orbits may differ in specific Keplerian parameters. However, they implement beam patterns that are as similar as possible to the required ones. In addition, they implement the required $(u,v)$ coverage, which is what was initially necessary.

\section*{CRediT authorship contribution statement}

\textbf{I. I. Bulygin}: Formal analysis, Methodology, Software, \textbf{M. A. Shchurov}: Conceptualization, Validation, Writing - Original Draft, \textbf{A. G. Rudnitskiy}: Resources, Validation, Writing - Original Draft.

\section*{Declaration of competing interest}

The authors declare that they have no conflict of interest concerning the publication of this manuscript.

\section*{Acknowledgements}

I. I. Bulygin acknowledges the financial support by the "BASIS" foundation and expresses its gratitude to the "Traektoria" foundation.

\appendix

\section{Kholshevnikov Metric}
\label{sect:appA}

When dealing with Keplerian orbital elements as a result of optimization, it is impossible to unambiguously determine during testing how close the resulting orbit is to the true one. To have a numerical criterion for this, we will use the Kholshevnikov metric \cite{Kholshevnikov2008, Kholshevnikov2020}. It is used to determine the orbit of the parent comet of meteor showers \cite{Kholshevnikov2016}.

Any Keplerian orbit is uniquely determined by the angular momentum vector $\mathbf{u}$ and the Laplace vector $\mathbf{v} \perp \mathbf{u}$. If we make them of the same dimension, the metric on a set of orbits can be defined as:
\begin{equation}
    d_{1,2} = \sqrt{(\mathbf{u}_1 - \mathbf{u}_2)^2 + (\mathbf{v}_1 - \mathbf{v}_2)^2}
\end{equation}

Where $\mathbf{u}$ and $\mathbf{v}$ are vectors co-directed with the angular momentum and the Laplace vector, normalized to the characteristic scale of the near-Earth orbits of the SVLBI -- the Earth's diameter $D_\oplus$:

\begin{equation}
    \begin{split}
        u_x &= \sqrt{p/D_\oplus} \sin i \sin \Omega \\
        u_y &= - \sqrt{p/D_\oplus} \sin i \cos \Omega \\
        u_z &= \sqrt{p/D_\oplus} \cos i
    \end{split}
\end{equation}

\begin{equation}
    \begin{split}
        v_x &= e \sqrt{p/D_\oplus} \left( \cos \omega \cos \Omega - \cos i \sin \omega \sin \Omega \right) \\
        v_y &= e \sqrt{p/D_\oplus} \left( \cos \omega \sin \Omega + \cos i \sin \omega \cos \Omega \right) \\
        v_z &= e \sqrt{p/D_\oplus} \sin i \sin \omega
    \end{split}
\end{equation}

Here $p = a\, (1 - e^2)$ is the focal parameter. However, these two vectors are related by the equation $\mathbf{u}\mathbf{v} = 0$, so five variables are sufficient to describe the problem instead of six. The calculations use the coordinates of the Laplace vector in a plane perpendicular to the vector $\mathbf{u} $, namely:

\begin{equation}
    \begin{split}
        v_1 &= v_x \sin \Omega \cos i - v_y \cos \Omega \cos i - v_z \sin i  \\
        v_2 &= v_x \cos \Omega  + v_y \sin \Omega
    \end{split}
\end{equation}

In one variant of the minimization process, the components of the vectors $(u_x, u_y, u_z, v_1, v_2)$ are used as variables. The reverse transition from these parameters to the Keplerian orbital elements occurs as follows:

\begin{equation}
    \begin{split}
        v_x &= v_1 \cos i \sin \Omega - v_2 \cos \Omega \\
        v_y &= - v_1 \cos i \cos \Omega + v_2 \sin \Omega \\
        v_z &= - v_1 \sin i
    \end{split}
\end{equation}

\begin{equation}
    \begin{split}
        e &= \dfrac{|\mathbf{v}|}{|\mathbf{u}|} \\
        a &= \dfrac{|\mathbf{u}|^2}{1 - e^2} \cdot D_\oplus \\
        i &= \arccos \frac{u_z}{|\mathbf{u}|} \\
        \Omega &= \mathtt{arctan2} \left(u_x, - u_y \right) \\
        \omega &= \mathtt{arctan2} \left(v_z, v_x \cos \Omega + v_y \sin \Omega \right) \\
    \end{split}
\end{equation}

\bibliographystyle{elsarticle-num} 
\bibliography{biblio}
\end{document}